\documentclass[a4paper,11pt]{article}  
\usepackage{amsmath,amssymb,fullpage}
\usepackage{algorithmic}

\newcommand{\conj}[1]{\ensuremath{{\overline{#1}}}}
\newcommand{\proof}{\noindent {\sc Proof:}~}
\newcommand{\foorp}{\hfill $\Box$}

\def\G{\ensuremath{\mathcal{G}}}
\def\N{\ensuremath{\mathbb{N}}}

\def\lm{\ensuremath{{\mathsf{lm}}}}
\def\lt{\ensuremath{{\sf lt}}}
\def\lc{\ensuremath{{\sf lc}}}
\def\l{\ensuremath{\langle}}
\def\r{\ensuremath{\rangle}}

\newtheorem{Theo}{Theorem}

\newtheorem{Cor}{Corollary}
\newtheorem{Prop}{Proposition}

\newtheorem{Not}{Notation}
\title{Structure of lexicographic Gr\"obner bases in three variables
of ideals of dimension zero}
\author{X.~Dahan\footnote{Supported by the GCOE program ``Math-for-Industry''
of Ky\^ush\^u university}\bigskip
\\
Dep\textsuperscript{t} of Mathematics, Ky\^ush\^u university, Japan\\
{\tt dahan@math.kyushu-u.ac.jp}}
\date{}
\begin{document}
\maketitle

\begin{abstract}
We generalize  the structural theorem of Lazard in 1985,
from 2 variables to 3 variables.
We use the Gianni-Kalkbrener result to do this,
which implies some restrictions
inside which lies the case of a radical ideal.
\end{abstract}

\section{Introduction}
Let $I$ be a zero-dimensional ideal of a polynomial
ring $R[x,y,z]$ over a N\oe therian domain $R$.
The {\em lexicographic order}
$\prec:=\prec_{lex(x,y,z)}$, for which
$x\prec y\prec z$,
is put on the monomials 
of $k[x,y,z]$ 
Given a polynomial $p\in k[x,y,z]$,
the {\em leading monomial} of $p$,
denoted $\lm_{\prec}(p)$ is the largest
monomial for $\prec$ occurring in $p$.
The coefficient in $R$ in front of $\lm_{\prec}(p)$
is called the {\em leading coefficient} of $p$,
denoted $\lc_{\prec}(p)$.
It might also be convenient to define the
{\em leading term} of $p$
denoted $\lt_{\prec}(p)$ equal to $\lc_{\prec}(p) \lm_{\prec}(p)$.

The {\em ideal of leading terms} of $I$ is the ideal
of $R[x,y,z]$ generated by the leading terms of elements
of $I$; it is equal to $\l \lt_{\prec}(I)\r$.
Since $R$ is N\oe therian, there is a finite set of generators
of this ideal.
A {\em Gr\"obner basis} of $I$
is a finite set of elements in $I$,
$g_1,\ldots,g_s$ such that $\l \lt_{\prec}(g_1),\ldots ,
\lt_{\prec}(g_s)\r = \l \lt_{\prec}(I)\r$.

In our case, we will take $R=k$ a field.
Note that then $\l \lt_{\prec}(I)\r$
is equal to $\l \lm_{\prec}(I)\r$.
This last ideal being a monomial ideal, it admits a minimal basis
of monomials $m_1,\ldots,m_s$; Then a Gr\"obner basis
$g_1,\ldots,g_s$ is {\em minimal} if $\lm_{\prec}(g_i)=m_i$ for all $i$.
It is  {\em monic} if $\lc_\prec(g_i)=1$ for all $i$.

From now on, the monomial order will always be assumed
to be $lex(x,y,z)$ and th symbol $\prec$ will be omitted
in $\lm_\prec,\lc_\prec$ and $\lt_\prec$.

\begin{Not}
Consider the rings $R_1:=k[x]$ and $R_2:=k[x,y]$.
Given $p \in k[x,y,z]=R_1[y,z]=R_2[z]$, let $\lc_1(p)\in R_1$
be the leading coefficient of $p$ for the lexicographic order $\prec_{lex(y,z)}$
on $R_1[y,z]$ and let $\lc_2(p)\in R_2$ be the leading coefficient
of $p\in R_2[z]$.

Furthermore, let $\lm_1(p)$ and $\lm_2(p)$ be the monomials
such that
$\lt(p)=\lc_1(p)\lm_1(p)=\lc_2(p)\lm_2(p)$.
\end{Not}
Moreover, we make the following  assumption:
\smallskip

\noindent {\bf Assumption:}
The ideal $I$ will be supposed {\em zero-dimensional},
or, equivalently the $k$-algebra $k[x,y,z]/I$ is supposed finite.
We are given a minimal and monic Gr\"obner basis 
$\G:=\{ g_1,\ldots,g_s\}$ of $I$,
indexed in a way that $\lm(g_1)\prec \lm(g_2) \prec \cdots \prec \lm(g_s)$.
 \smallskip

We recall some basic facts about the Gr\"obner basis $\G$: 
\begin{itemize}
\item  $g_1\in k[x]$ and $\lm(g_s)=z^{d_s}$ for some $d_s\in \N^\star$
(we say that $\lm(g_s)$ is {\em pure power} of $z$).
\item Moreover, there exists $1<\ell(2)<s$ such that:
$\lm(g_{\ell(2)})=y^{d_{\ell(2)}}$ is a pure power of $y$ and
such that
$g_i \in k[x,y] \setminus k[x]$ for $1<i<\ell(2)$;
and $g_i\in k[x,y,z] \setminus k[x,y]$ for $i>\ell(2)$.
\item {\bf Elimination property:} the set of polynomials
$g_1 , \ldots , g_{\ell(2)}$ is a minimal lexicographic
Gr\"obner basis of the zero-dimensional ideal $I\cap k[x,y]$.
\end{itemize}

In 1985, Lazard in~\cite{Laz85}  proves the following.
\begin{Theo}[D. Lazard]
Let $J\subset k[x,y]$ be a zero-dimensional ideal,
and $f_1,\ldots,f_r$ a minimal lexicographic Gr\"obner basis
of $I$ for $x\prec_{lex(x,y)} y$. Then:
$$
\lc_1(f_i)\in k[x_1]  \quad \text{divides} \quad \lc_1(f_j) \text{ for all }
\quad i \ge j, \qquad \text{and} \quad \lc_1(f_i)\quad \text{divides}
\quad \quad f_i \quad \text{as well.}
$$
\end{Theo}
It follows easily a factorization property of the polynomials
in such a Gr\"obner basis~\cite[Theorem~1~(i)]{Laz85}.
However, the formulation above
is more compact and handy, and is equivalent.
The main result of this paper is the following analogue
in the case of 3 variables:
\begin{Theo}\label{th:main}
Let $I$, $\G:=\{g_1,\ldots,g_s\}$ and  $\ell(2)$ be defined as above.
Then, for all $1\le j \le i\le s$ such that the variable $z$
appears in the monomials $\lm(g_i)$ and $\lm(g_j)$ with the {\em same
exponent}, holds:
$$\lc_1(g_i) \quad \text{divides} \quad \lc_1(g_j),\quad
\text{and if $I$ is radical:}\qquad \lc_1(g_i)\quad \text{divides}\quad
g_i \quad \text{as well}.$$
Furthermore, in the later case, for all $i >\ell(2)$, $g_i\in \l \lc_2(g_i) ,
g_1\r$.  
\end{Theo}
The proof will occupy the next section.
There is one corollary to this theorem
in the context of ``stability of Gr\"obner bases under specialization'',
which generalizes the theorem of Gianni-Kalkbrener~\cite{Gi87,Ka87},
and improves the theorem of Becker~\cite{Be94}
(but  holds only with 3 variables).
\begin{Cor}\label{cor:main}
Let us assume $I$  radical.
Let $\alpha$ be a root of $g_1$,
$\phi:\conj{k}[x,y,z]\rightarrow\conj{k}[y,z]$, $x\mapsto \alpha$,
and $g\not=g_1$ a polynomial
among the Gr\"obner basis.
Then, either $\phi(g)=g(\alpha,x,z)=0$, or $\phi(\lc_1(g))\not=0$.
This implies that: $\lt(\phi(g))=\phi(\lt(g))$,
and in particular, that $\phi(\G)$ is a Gr\"obner basis.
\end{Cor}
\proof By Theorem~\ref{th:main},
we can write $g= \lc_1(g) A$ with $A=\frac g{\lc_1(g)}\in k[x,y,z]$.
Hence, if $\phi(\lc_1(g))=0$, then $\phi(g)=0$.
Else, since $\lt(A)=y^\bullet z^\bullet$, we get
$\phi(\lt(A))=\lt(\phi(A))$. But $\lt(g) = \lc_1(g) \lt(A)$,
from which follows $\phi ( \lt(g) ) =  \phi(\lc_1(g))
\phi(\lt (A))$.
On the other hand, $\lt(\phi(g))=\lt (\phi(\lc_1(g)) \phi(A))=\phi(\lc_1(g)) \lt(\phi(A))$.
\foorp
\medskip

Gianni-Kalkbrener's result~\cite{Gi87,Ka87} concerns the easier
case where all the variables but the largest one for $\prec$ are specialized.
\smallskip

\noindent {\bf Gianni-Kalkbrener.}
The map $\phi$ is therein $\phi : \conj{k}[x,y,z]\rightarrow
\conj{k}[z]$, $x,y\mapsto\alpha,\beta$
 for $(\alpha,\beta)$
a solution of the system $g_1,\ldots,g_{\ell(2)}\subset k[x,y]$.
For any $g$ in the Gr\"obner basis $\G$
such that $g\in k[x,y,z]\setminus k[x,y]$,
they show that either $\phi(g)=0$ or
$\deg_z(\lt(\phi(g))) = \deg_z(\phi(\lt(g)))$,
which implies
$\phi(\lt(g)) = \lt(\phi(g))$.
\smallskip

Becker~\cite{Be94} has generalized partly this result to the case
of a map $\phi$ that specializes the $t$ lowest variables
for $\prec$. Taking $t=1$, this covers the case of Corollary~\ref{cor:main},
but is weaker: it does also say that $\phi(\G)$ remains a Gr\"obner basis,
while assuming that for $g\in \G$,
$\phi(\lt(g))$ may be a term with a monomial
strictly smaller for $\prec$ than the monomial in the term $\lt(\phi(g))$ (see the definition
of the integer $r'$ during the proof of Prop.~1  page~4
of~\cite{Be94}.
With the notations on the same page of~\cite{Be94} we see $r'<r$;
Corollary~\ref{cor:main} above implies $r=r'$).
It can not be said that: $\phi(\lt(\G)) =\lt (\phi (\G))$. 

Concerning  previous works, let us mention that Kalkbrener~\cite{Ka97}
has expanded Becker's result to the more general elimination monomial
orders. Still, staying in the purely lexicographic case,
it does not enhance  the theorem of Becker.

\section{Proof of Theorem~\ref{th:main}}
The main ingredient of the proof consists in generalizing two lemmas of Lazard.
These refers to Lemma~2, and Lemma~3 of~\cite{Laz85}.
We shall explain that a weaker form holds with a larger
number of variables.
The version of interest here concerns the case of 3 variables.
It is nonetheless easy to produce a version with an arbitrary number
of variables.
Let us first introduce some notations
for exponents:
\begin{Not}
Let $f\in k[x,y,z]$ non zero, with leading monomial
$\lm(f)=x^ay^bz^c$. The 3 notations $\alpha_x(f)$,
$\alpha_y(f)$ and $\alpha_z(f)$ will denote
$a,b$ and $c$ respectively.

If $g_i$ is among the Gr\"obner basis $\G=\{g_1,\ldots,g_s\}$,
the shortcuts $\alpha_x(i),\alpha_y(i),\alpha_z(i)$
will be used instead of $\alpha_x(g_i),\alpha_y(g_i),\alpha_z(g_i)$
\end{Not}

\begin{Prop}\label{prop:1}
Let $1\le j < i \le s$ be such that
$\alpha_y(j) \le \alpha_y(i)$
and $\alpha_z(j)\le \alpha_z(i)$.
Then $\lc_1(g_i)$ divides $\lc_1(g_j)$.
\end{Prop}
\proof
Let $a:=g_j y^{\alpha_y(i)-\alpha_y(j)} z^{\alpha_z(i)-\alpha_z(j)}$.
The multivariate division algorithm with respect
to $\prec$ of $a$ by $[g_i]$ gives:
$$
a = q g_i + r,\quad \text{with} \
q\not=0 \Rightarrow \lm(a)\preccurlyeq \lm(q g_i),
$$
and $lm(g_i)$ does not
divide any monomial occurring in $r$.

By definition of $a$, $\lm(g_i)\, |\, \lm(a)$
so that $q\not=0$, hence $\lm(q g_i)\preccurlyeq \lm(a)$
holds:
$$
\lm(q g_i) = \lm(q) x^{\alpha_x(i)} y^{\alpha_y(i)}
z^{\alpha_z(i)} \preccurlyeq 
x^{\alpha_x(j)} y^{\alpha_y(i)}
z^{\alpha_z(i)}=\lm(a)\quad
\Rightarrow 
\lm(q) x^{\alpha_x(i)} \preccurlyeq x^{\alpha_x(j)}.
$$
By an elementary property of the lexicographic order $\prec_{lex(x,y,z)}$,
this implies $\lm(q)\in k[x]$ and therefore
$q\in k[x]$.
Next, the equality $r = a - q g_i$ gives:
$$
\lm(r) = \lm(a - q g_i)
\preccurlyeq \max\{ \lm(a) ; \lm(q g_i)\} =
x^{\max\{\alpha_x(q g_i),\alpha_x(a)\}}
y^{\alpha_y(i)}z^{\alpha_z(i)}.
$$
Again, property of lexicographic order
implies $\alpha_z(r)\le \alpha_z(i)$
and if $\alpha_z(r)=\alpha_z(i)$
then $\alpha_y(r) \le \alpha_y(i)$.
We distinguish three cases;
in the first two ones the conclusion
of the theorem holds, and the third case
never happens.
\smallskip

{\em Case 1:} $\alpha_z(r) < \alpha_z(i)$.
Then $\lc_1(a)=q \lc_1(g_i)$, and $\lc_1(a)=\lc_1(g_j)$,
this concludes the proof. 
\smallskip

{\em Case 2:} Else $\alpha_z(r) = \alpha_z(i)$,
and $\alpha_y(r)<\alpha_z(i)$. Similarly,
this shows that $\lc_1(a)=q\lc_1(g_I)$,
concluding the proof.
\smallskip

{\em Case 3:} Else $\alpha_z(r)=\alpha_z(i)$
and $\alpha_y(r) = \alpha_y(i)$.
Since $\lm(g_i) \nmid \lm(r)$,
necessarily $\alpha_x(i) > \alpha_x(r)$.
On the other hand, $r\in \l g_j,g_i\r\subset I$
implies that there exists $1\le k\le s$
such that $\lm(g_k)\, |\, \lm(r)$.
Therefore,
$\alpha_x(k)\le \alpha_x(r)<\alpha_x(i)$,
and in this case $\alpha_y(k) \le \alpha_y(r)=\alpha_y(i)$,
$\alpha_z(k)\le \alpha_z(r)=\alpha_z(i)$.
This means $\lm(g_k)\, |\, \lm(g_i)$,
and $i\not=k$, which is impossible since 
the Gr\"obner basis is minimal. \foorp

\begin{Prop}\label{prop:2}
For any $i>1$, the polynomial  $g_i$ of the
 the Gr\"obner basis $\G$ verifies: $\lc_1(g_i)$ divides $\lc_2(g_i)$.
\end{Prop}
\proof
Define,
$$e_i:=\max\{ \alpha_y(\ell)\ \text{ s.t }
\ \alpha_y(\ell) < \alpha_y(i),\ \alpha_z(\ell) \le
\alpha_z(i)\}
\quad \text{and}\quad j:=\max\{\ell <i\ \text{ s.t }
\ \alpha_y(\ell) = e_i\} 
$$
Note that $e_i$ is well-defined
because $i>1$ and $\alpha_y(1)=\alpha_y(g_1)=0$.
This also shows that $j$ is well-defined.
By Proposition~\ref{prop:1}, $\lc_1(g_i)$
divides $\lc_1(g_j)$. Let
$$
a:= \frac{\lc_1(g_j)}{\lc_1(g_i)} g_i,
\quad \text{and} \quad b:= a - g_j y^{\alpha_y(i)-\alpha_y(j)}
z^{\alpha_z(i)-\alpha_z(j)}.
$$
By construction, $\lm(b)\prec y^{\alpha_y(i)}
z^{\alpha_z(i)}$. Furthermore, $b\in \l g_i,g_j\r\subset I$
so its normal form modulo the Gr\"obner basis
of $I$ is 0. The multivariate division equality
with respect to $\prec$ of $b$ by $[g_1,\ldots,g_s]$
is written: $b=\sum_{1\le \ell\le s}b_\ell g_\ell$.
If $b_\ell \not = 0$, then $\lm(b_\ell g_\ell)
\preccurlyeq \lm(b)\prec y^{\alpha_y(i)}
z^{\alpha_z(i)}$. The inequality $\lm(b_\ell) \prec 
y^{\alpha_y(i)}
z^{\alpha_z(i)}$ follows, which is possible 
only if $\ell \le i-1$.
Otherly said, $b=\sum_{1\le \ell \le i-1} b_\ell g_\ell$.

It follows that
$
a= \sum_{\ell\not=j \atop \ell=1}^{i-1} 
b_\ell g_\ell + g_j (b_j + y^{\alpha_y(i)-\alpha_y(j)}
z^{\alpha_z(i)-\alpha_z(j)})$,
and that:
\begin{equation}\label{eq:lc2a}
\lc_2(a) = \sum_{\ell\not=j \atop \alpha_z(g_\ell b_\ell)
=\alpha_z(i)} 
\lc_2(b_\ell)\lc_2( g_\ell) + \lc_2(g_j) (\epsilon \lc_2(b_j) + 1),
\end{equation}
with $\epsilon =1$  if 
$\alpha_z(j)+\alpha_z(b_j)= \alpha_z(i)$
and
$\epsilon =0$  if $\alpha_z(j)+\alpha_z(b_j)\prec \alpha_z(i)$.
However $\lm(b_\ell g_\ell) \prec
y^{\alpha_y(i)} z^{\alpha_y(i)} $
and $\alpha_z(b_\ell g_\ell)=\alpha_z(i)$
imply that $\alpha_y(b_\ell) + \alpha_y(\ell) < \alpha_y(i)$.
In particular $\alpha_y(\ell)< \alpha_y(i)$
and consequently $\alpha_y(\ell)\le e_i$.
By definition of $j$,
this gives:
$\ell \le j$.
Proposition~\ref{prop:1} then yields: $\lc_1(g_j) \, |\,  \lc_1(g_\ell)$.

To conclude, note that Lazard's Lemma~4
in~\cite{Laz85}
proves that Prop.~\ref{prop:2} is true for
$1 < i \le \ell(2)$. So we can proceed by induction
on $i$ and assume that $\lc_1(g_\ell) \, |\,  \lc_2(g_\ell)$
for $2 \le \ell <i$. Applied in Equation~\eqref{eq:lc2a}:
$$
\lc_2(a) = \sum_{\ell \not=j\atop \alpha_z(g_\ell b_\ell)
=\alpha_z(i)}
\lc_2(b_\ell) \frac{\lc_2(g_\ell)}{\lc_1(g_\ell)}
\frac{\lc_1(g_\ell)}{\lc_1(g_j)} \lc_1(g_j)
+ \frac{\lc_2(g_j)}{\lc_1(g_j)}\lc_1(g_j)
(\epsilon \lc_2(b_j) + 1)\ \ \in\  k[x,y]
$$
Finally, $\frac{\lc_2(a)}{\lc_2(g_j)} =
\frac{\lc_2(g_i)}{\lc_1(g_i)} \in k[x,y]$.
\foorp
\medskip

This proves the first part of Theorem~\ref{th:main}.
The second part is based upon the previous proposition and the theorem
of Gianni-Kalkbrener.
The use of the later requires a restriction:
\begin{Prop}\label{prop:3}
Suppose there is an $1 \le i<s$
such that: $\lc_1(g_i)\not=1$,
there is a root $\alpha$  of $\lc_1(g_i)$
which is not a root of $\lc_1(g_{i+1})$.
Then, $g(\alpha,y,z)=0$
and $g_{i+1} \in \l x-\alpha , \lc_2(g_{i+1})(\alpha,y)\r$.
\end{Prop}
\proof Since $\lc_1(g_i)(\alpha)=0$, by Proposition~\ref{prop:2},
$\lc_2(g_i)(\alpha,y)=0$ as well. By Gianni-Kalkbrener,
this implies $g_i(\alpha,y,z)=0$.
Furthermore,  $\lc_1(g_{i+1})(\alpha)\not=0$,
implying $p_{\alpha}(y):=\lc_2(g_{i+1})(\alpha,y)\in \conj{k}[y]$ is not zero. 
Let $\beta \in \conj{k}$ be a root of this polynomial.
By Gianni-Kalkbrener, $g_{i+1}(\alpha,\beta,z)=0$,
showing that $g_{i+1}\in \l x-\alpha,p_{\alpha}\r$.
\foorp
\medskip

Note that if $I$ is radical, all elements $g_i$
for which $\lc_1(g_i)\not=1$ verify the assumption
on the root $\alpha$
of  Proposition~\ref{prop:3}. By an elementary use
of the Chinese remaindering theorem, we get the more general,
$g_{i+1} \in \l g_1,\lc_2(g_{i+1})\r$.
This proves the last part of Theorem~\ref{th:main}.

\section*{Conclusion}

It is likely that Theorem~\ref{th:main}
holds without the assumption $I$ radical.
This assumption was set to allow the use
of Gianni-Kalkbrener's result. A proof
circumventing it must be found.
Also, some experiments shown  that the results presented
here are certainly true in the case of more than 3 variables.


\bibliographystyle{plain}
\bibliography{/Users/dahan/main}
\end{document}